\documentclass[conference, 10pt]{IEEEtran}
\IEEEoverridecommandlockouts
\usepackage{cite}
\usepackage{amsmath,amssymb,amsfonts}
\usepackage{algorithmic}
\usepackage{graphicx}
\usepackage{textcomp}
\usepackage{xcolor}
\def\BibTeX{{\rm B\kern-.05em{\sc i\kern-.025em b}\kern-.08em
    T\kern-.1667em\lower.7ex\hbox{E}\kern-.125emX}}

\usepackage{mathtools}
\input{mysymbol.sty}

\usepackage{amsthm}
\usepackage{hyperref} 

\usepackage{algorithmic}
\usepackage{algorithm}
\usepackage{listings}
\usepackage{enumerate}
\usepackage{enumitem}
\usepackage{subcaption}
\usepackage{bbm} 

\date{\today}

\def\E{\mathbb{E}}

\usepackage{caption}
\begin{document}

\title{
Learning to Slice Wi-Fi Networks: A State-Augmented Primal-Dual Approach
\thanks{This work has been supported by NSF-Simons MoDL, Award 2031985, NSF AI Institutes program.}

}

\author{\IEEEauthorblockN{Yi\u{g}it~Berkay~Uslu}
\IEEEauthorblockA{
\textit{University of Pennsylvania}\\
Philadelphia, PA, USA \\
ybuslu@seas.upenn.edu}
\and
\IEEEauthorblockN{Roya Doostnejad}
\IEEEauthorblockA{
\textit{Intel Corporation}\\
Santa Clara, CA, USA \\
roya.doostnejad@intel.com}
\and
\IEEEauthorblockN{Alejandro Ribeiro}
\IEEEauthorblockA{
\textit{University of Pennsylvania}\\
Philadelphia, PA, USA \\
aribeiro@seas.upenn.edu}
\and
\IEEEauthorblockN{Navid NaderiAlizadeh}
\IEEEauthorblockA{
\textit{Duke University}\\
Durham, NC, USA \\
navid.naderi@duke.edu}
}

\maketitle

\begin{abstract}
Network slicing is a key feature in 5G/NG cellular networks that creates customized slices for different service types with various quality-of-service (QoS) requirements, which can achieve service differentiation and guarantee service-level agreement (SLA) for each service type. In Wi-Fi networks, there is limited prior work on slicing, and a potential solution is based on a multi-tenant architecture on a single access point (AP) that dedicates different channels to different slices. In this paper, we define a flexible, constrained learning framework to enable slicing in Wi-Fi networks subject to QoS requirements. We specifically propose an unsupervised learning-based network slicing method that leverages a
state-augmented primal-dual algorithm, where 
a neural network policy is trained offline to optimize a Lagrangian function and the dual variable dynamics are updated online in the execution phase. We show that state augmentation is crucial for generating slicing decisions that meet the ergodic QoS requirements. 
\end{abstract}

\begin{IEEEkeywords}
Wi-Fi slicing, 5G, QoS, constrained learning, Lagrangian duality, primal-dual, state augmentation.
\end{IEEEkeywords}

\maketitle

\section{Introduction}\label{sec:intro}
In enterprise settings, it is vital to manage network operations to support multiple use cases with
different requirements \cite{shen2020aiassisted, wu2022mobile6g}. 
In addition, 3GPP includes architectures to integrate Wi-Fi in converged connectivity (5G + Wi-Fi) in enterprise \cite{lemes2022tutorial}. Network slicing allows an access point (AP) to allocate the network resources across service-level agreement (SLA) categories rather than individually across many users/flows. Defining dynamic slicing in Wi-Fi with guaranteed quality-of-service (QoS), in a similar fashion as 5G, has the potential to provide mechanisms for seamless dynamic traffic steering across licensed and unlicensed networks. Current Wi-Fi slicing solutions are based on a multi-tenant architecture on a single AP by installing different SSIDs \cite{nerini20215g, networkslicing2018}. 
Despite being simple to implement, this method has drawbacks, including the increased overhead for beacon transmission and probe responses by virtual APs, the lack of QoS differentiation inside each SSID slice, and the limitation of association to one slice.

Machine learning approaches are ubiquitous in resource allocation/optimization problems in wireless networks, and there has been growing attention to tackling network slicing problems through learning-based approaches, with reinforcement learning (RL) methods being the most common \cite{zangooei2023reinforcement, yang2023advancing, liu2020constrained}. However, prior work analyzing network slicing with constraints imposed by QoS requirements is relatively scant. For example, in \cite{yang2023advancing, liu2020constrained}, the QoS requirements are taken into account by either (i) directly adding the weighted QoS requirements to the objective, or (ii) using special functions such as log-barrier functions and an unconstrained, penalized aggregate reward, which is optimized using conventional RL algorithms. These approaches generally lack feasibility guarantees---which is partially remedied by projections to feasible policies---and are highly sensitive to suitable selection of penalty weights. 

Constraint-aware online deep reinforcement learning (DRL) training is explored in \cite{liu2021onslicing} by leveraging Lagrangian primal-dual methods with a proactive baseline switching. Similarly, \cite{agostini2022learning} handles the QoS constraints by reformulating the learning problem in the dual domain, which is then solved via a primal-dual policy gradient algorithm. The primal-dual methods, although effective, do not guarantee convergence to an optimal policy similar to penalized methods.

In this paper, we define a constrained learning framework to enable dynamic network slicing in Wi-Fi. Sections~\ref{sec:network_slicing} and~\ref{sec:System Model} introduce a flexible Wi-Fi network slicing framework and the problem setup. Departing from the aforementioned works, we formulate the network slicing problem as a constrained radio resource management (RRM) optimization problem and develop an unsupervised state-augmented primal-dual algorithm in Section~\ref{sec:problem formulation}. The notion of state augmentation---proposed in \cite{calvo2021state} for constrained RL and in \cite{StateAugmented_RRM_GNN_naderializadeh_TSP2022} for wireless resource allocation---incorporates the Lagrangian dual multipliers into the state space similar to a closed-loop feedback system. The novelty of our solution lies in that state-augmentation allows for training a single model to learn Lagrangian minimizing policies over a range of dual multipliers and running dual updates online to sample from different learned policies. This approach results in a practical online algorithm that samples from an optimal (stochastic) policy and exhibits feasibility and near-optimality guarantees, which do not hold for the regularized and conventional primal-dual methods. We demonstrate these attributes of state augmentation in the context of network slicing with ergodic throughput and latency QoS requirements through numerical experiments in Section~\ref{sec:experimental results}.

\section{Network Slicing in Wi-Fi}\label{sec:network_slicing}
A flexible framework is defined to enable slicing in Wi-Fi, allocating resources to assure SLAs, including maximum delay for low-latency (L) and minimum data rate/throughput for high-throughput (H) cases, while maximizing resource units (RU) available to other slices including best-effort (B). The learning-based slicing module collects data from the ongoing traffic and creates, modifies, and removes slices accordingly so that there is minimal impact on traffic and services in other network slices in the same network. 

\subsection{Slice ID}
Each slice is uniquely identified by its Network Slice ID or Slice type indicator. Both the devices and the network may initiate slice configuration. The device may request to set up a new connection, but the infrastructure may accept or reject the request based on channel conditions, QoS requirements, and available network capacity. The network may also validate if the user has the right to access a network slice. In case the network initiates slice configuration, the device will convey the provided slice ID.

\subsection{Resource Isolation and Management}
A key enabler for slicing is the ability to isolate traffic between different slices, for which there are two solutions.
\subsubsection{Scheduling Resources in OFDMA}
 Each slice is dynamically configured and allocated a certain portion of the frequency band. Applications in the same slice undergo transmission scheduling and are allocated RUs from the assigned bandwidth. In uplink, a trigger-based approach is used to schedule applications in each slice. A minimum RU is allocated to all applications to support legacy devices.
\subsubsection{Multi-Link Operation (MLO)}
 The MLO process, newly defined in IEEE802.11be, may facilitate device association to all slices and seamless movement across slices. An MLO device may simultaneously connect to multiple slices and support different traffics with different SLAs.
 
The slicing framework associates each slice with a main link and one or more auxiliary links, in which the slice has a lower priority with respect to the slices that use the link as the main. The slices that
have a low priority on a link will be served in that link only if the flows that use it as primary do not occupy
all the available resources. This makes the cross-link slicing mechanism adaptive to the load of each slice. The concept of main and auxiliary multi-link for slicing can also be extended to include licensed bands for multi-spectral slicing.
\section{Network Slicing Model}\label{sec:System Model}
Consider a wireless network where each access point (AP) is serving $| \ccalU |$ flows, indexed by $i \in \ccalU$, over a $W$ MHz channel. Each flow $i \in \ccalU$ receives data from the application at an average rate $\bar{\mu}_i > 0$, which must be transmitted from the AP to the user device. The performance requirements each flow must meet are dictated by its SLA, and we consider three SLA categories with distinct ergodic QoS requirements: 
(i) \emph{high-throughput (H)}, (ii) \emph{low-latency (L)}, and (iii) \emph{best-effort (B)}. 
The high-throughput SLA requirement is a minimum throughput of $\rminht$ bps/Hz, while the low-latency SLA requirement is $\lmaxll$ milliseconds of maximum packet delay/latency. We specify no QoS target for the best-effort SLAs.
We assume each flow is associated with only one SLA and denote by $\Uht$, $\Ull$, and $\Ube$ the disjoint subsets of flows with high-throughput, low-latency, and best-effort SLAs, respectively. 

The network operates over $\Tslice$ discrete epochs, indexed by $\tslice = 0, \ldots, \Tslice-1$. We refer to the discrete epoch $\tslice$ as the $\tslice$th slicing window. Each slicing window lasts $\tau_{\max}$ seconds
and we index any timestamp within that slicing window with $\tau \in [0, \tau_{\max})$. Each flow $i$ maintains a queue of size $Q^{\tslice}_i(\tau) \in [0, Q_{\max}]$ that evolves with a dynamic traffic arrival rate $\alpha^{\tslice}_i(\tau) \sim \alpha(\mu^{\tslice}_i)$ and a dynamic transmission rate $r^{\tslice}_i(\tau)$.

\begin{figure}[t!]
    \centering
    \includegraphics[width = .9\linewidth]{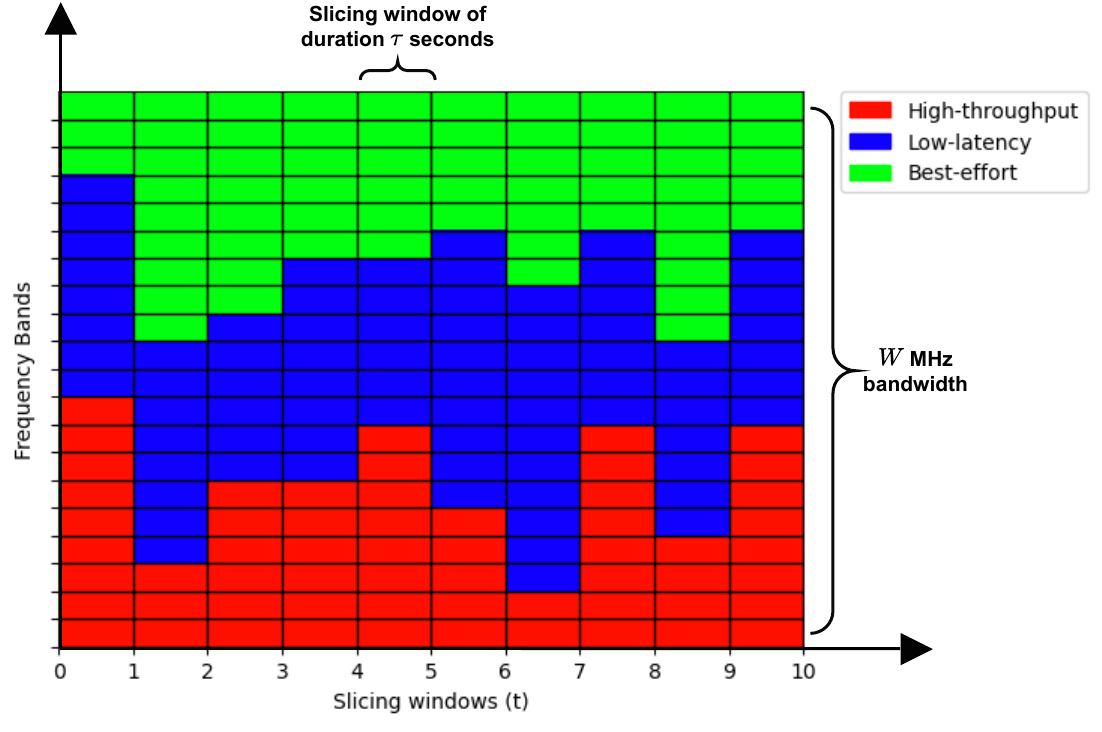}
    \caption{A network slicing diagram. Frequency subbands are allocated to three types of traffic in each slicing window.}
    \label{fig:network-slicing-diagram}
\end{figure}%

The QoS metrics throughput and latency for a flow $i \in \ccalU$ over the $\tslice$th slicing window, denoted by $r_i(\tslice)$ and $\ell_i(\tslice)$, respectively, are determined primarily by the dynamic rate $r^{\tslice}_i(\tau)$ which itself is determined by the frequency resource units (FRUs) allocated to flow $i$ by the AP and the channel state which is an indicator of the signal-to-interference-plus-noise ratio (SINR) experienced by the device receiving flow $i$. The channel state 
varies randomly from each slicing window to window, e.g., following a Rayleigh fading channel model.

The goal of a slicing policy is to allocate the FRU resources across SLA categories while ensuring the QoS requirements are satisfied. We represent a slicing decision/policy as
\begin{align}\label{eq:slicing_policy}
    \bbp(\tslice) = \big( \pht(\tslice), \, \pll(\tslice), \, \pbe(\tslice) \big)^\top, \quad \tslice = 0, \ldots, \Tslice-1,
\end{align}
where the entries of $\bbp(\tslice)$ denote the fraction of the total bandwidth allocated to the respective SLA category at the beginning of the slicing window $t$ and all flows associated with the same SLA category access the same slice of resources. For example, if $\bbp(\tslice) = (1/2, 1/3, 1/6)$ and $W = 20$ MHz, then any high-throughput flow has access to $(1/2) \cdot 20 = 10$ MHz bandwidth. Within each slicing window, the AP further allocates sub-bands at a faster time scale according to some given scheduling algorithm (e.g., round-robin, max-SINR, etc.). Note that the slicing decisions are updated at the beginning of each slicing window. See Fig.~\ref{fig:network-slicing-diagram} for an illustrative diagram. 

Due to the dynamic and complex nature of the network traffic, wireless channel conditions, and so on, the network slicing decisions should occur opportunistically across slicing windows and learning-based optimization techniques can be leveraged. Let $\bbs_{\tslice}\in \mathcal{S}$ be the overall network state at the beginning of the slicing window $\tslice$ and $\bbtheta \in \bbTheta \subset \mathbb{R}^d$ be a finite-dimensional (e.g., neural-network) parametrization. 

We define the network slicing policy as
\begin{equation}\label{eq_resource_allocation}
    \bbp(\tslice) = \bbp^{\bbtheta}(\bbs_{\tslice}; \bbtheta)
\end{equation}
that takes the network state $\bbs_{\tslice}$ as input and outputs the corresponding slicing decisions. We include the parametrization $\bbtheta$ in the superscript to avoid confusion when we introduce a state-augmented parametrization in the next section. Our goal is to optimize the parameter vector $\bbtheta$ to maximize the system performance subject to the QoS requirements and the total bandwidth budget of $\pht(\tslice) + \pll(\tslice) + \pbe(\tslice) \leq 1$. Note that we can drop the total bandwidth budget constraint in subsequent derivations since it holds with equality for any optimal slicing decision and can be implicitly satisfied by the design of the learning architecture (e.g., a softmax layer).

We define the network-slicing optimization problem as
\begin{equation}
\begin{aligned}\label{eq:optimization_problem}
    \min_{ \bbtheta \in \bbTheta } \quad& \frac{1}{\Tslice} \sum_{\tslice = 0}^{\Tslice-1} f_0\big(\bbs_{\tslice}, \bbp^{\bbtheta}(\bbs_{\tslice}; \bbtheta) \big) \\
    \text{subject to} \quad& \frac{1}{\Tslice} \sum_{\tslice = 0}^{\Tslice-1} \bbf \big(\bbs_{\tslice}, \bbp^{\bbtheta}(\bbs_{\tslice}; \bbtheta) \big) \leq 0,%
\end{aligned}
\end{equation}
where 
$f_0: \mathcal{S} \times \mathbb{R}^3 \to \mathbb{R}$ is the objective (e.g., negated total network throughput), $\bbf: \mathcal{S} \times \mathbb{R}^3 \to \mathbb{R}^c$ is the vector-valued constraint function (e.g., minimum throughput and maximum packet latency QoS) for $c$ constraints in total. Note that the objective and the constraints are derived based on the ergodic averages of the respective metrics. 

The goal of this paper is to develop a learning algorithm to solve~\eqref{eq:optimization_problem} for any given network configuration and sequence of network states $\{\bbs_{\tslice} \}_{\tslice = 0}^{\Tslice-1} \in \ccalS$.

\section{Proposed State-Augmented Algorithm}\label{sec:problem formulation}
We move to the Lagrangian dual domain and write the Lagrangian function for the parametrized problem as
\begin{align}
\ccalL(\bbtheta, \bblambda) = \frac{1}{\Tslice} \sum_{\tslice = 0}^{\Tslice-1}  \Big[ &f_0\big(\bbs_{\tslice}, \bbp^{\bbtheta}(\bbs_{\tslice}; \bbtheta) \big) + \bblambda^\top \bbf \big(\bbs_{\tslice}, \bbp^{\bbtheta}(\bbs_{\tslice}; \bbtheta) \big) \Big], \label{eq:Lagrangian_parametrized}
\end{align}
where $\bblambda = (\lambda_1, \ldots, \lambda_c)^\top \in \mathbb{R}^c_{+}$ is the vector of nonnegative dual multipliers corresponding to the QoS constraints. Defining the dual function as $d(\bblambda) := \min_{\bbtheta \in \bbTheta} \ccalL(\bbtheta, \bblambda)$, the primal and dual problems are related by
\begin{align}
    D^{\star}_{\theta} := \max_{\bblambda \in \mathbb{R}^c_{+}} d(\bblambda) &= \max_{\bblambda \in \mathbb{R}^c_{+}} \min_{\bbtheta \in \bbTheta} \ccalL(\bbtheta, \bblambda) \leq P^{\star}_{\theta} \label{eq:dual_problem_parametrized}
\end{align}
where $P^{\star}_{\theta}$ is the optimum value of parametrized primal problem and $P^{\star}_{\theta} - D^{\star}_{\theta}$ is the duality gap. While the duality gap is generally nonzero for the parametrized problem, it becomes sufficiently small as the parametrization becomes richer~\cite{paternain2019zerodualitygap}. More importantly, although both the Lagrangian in \eqref{eq:Lagrangian_parametrized} and the dual problem \eqref{eq:dual_problem_parametrized} are nonconvex in $\bbtheta$, the outer maximization in the latter is a convex problem. In practice, the inner exact minimization corresponding to the primal policy optimization is also replaced by stochastic gradient-descent 
which results in an iterative primal-dual algorithm given by
\begin{equation}\label{eq:DGA}
\begin{aligned}
    \bbtheta_{k+1} &= \bbtheta_{k} - \eta_{\bbtheta} \nabla_{\bbtheta} \ccalL(\bbtheta_k, \bblambda_k), \\
    \bblambda_{k+1} &= \Big[ \bblambda_k + \eta_{\bblambda} \bbf(\bbtheta_{k+1}) \Big]_{+},
\end{aligned}
\end{equation}
where $k = 0, 1, \ldots,$ is an iteration index, $\eta_{\bbtheta}$ and $\eta_{\bblambda}$ are the step sizes for the primal and dual variables, respectively, and $[\cdot]_+ := \max(\cdot, \bb0)$. The trajectories $(\bbtheta_k, \bblambda_k)_{k \geq 0}$ generated by running \eqref{eq:DGA} lead to near-optimal and feasible decisions asymptotically under some mild assumptions \cite{calvo2021state, StateAugmented_RRM_GNN_naderializadeh_TSP2022}.

\begin{figure*}[t!]
\centering
    \begin{subfigure}[t]{.49\linewidth}
    \centering
    \includegraphics[width = \linewidth]{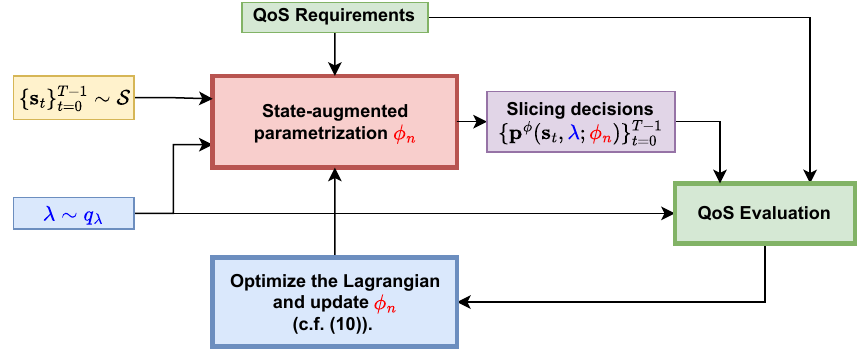}
    \subcaption{Training phase}
    \end{subfigure}\hfill%
    \begin{subfigure}[t]{.49\linewidth}
    \centering
    \includegraphics[width = \linewidth]{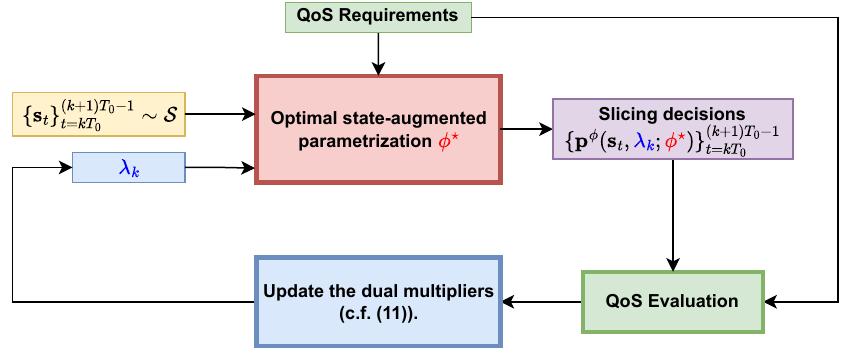}
    \subcaption{Execution phase}
    \end{subfigure}\hfill%
\caption{(a) Offline training, and (b) online execution phases of the proposed state-augmented slicing algorithm.}
\label{fig:train-and-test}
\end{figure*}%

Running primal-dual algorithms online in practice is difficult and stopping the offline training at an arbitrary iteration $k$ does not necessarily guarantee that $\bbtheta_k$ yields a feasible slicing policy. In fact, cycloscillatory behavior of constraints is a common phenomenon. To remedy these challenges, we resort to a state-augmented parametrization $\bbphi \in \bbPhi$, proposed in \cite{calvo2021state, StateAugmented_RRM_GNN_naderializadeh_TSP2022}, where the network state $\bbs$ is \emph{augmented} with the dual multiplier vector $\bblambda$ corresponding to the QoS constraints at the slicing window $\tslice$ and we denote the slicing policy as $\bbp^{\bbphi}(\bbs, \bblambda; \bbphi)$. This way, the dual multipliers, which track the evolution of constraint satisfactions/violations over time, are incorporated into the state space. We define the Lagrangian
\begin{equation}
\begin{aligned}
  \ccalL_{\bblambda}(\bbphi) = \frac{1}{\Tslice} \sum_{\tslice = 0}^{\Tslice-1}  \Big[ &f_0 \big(\bbs_{\tslice}, \bbp^{\bbphi}(\bbs_{\tslice}, \bblambda; \bbphi) \big) \\
  &+ \bblambda^\top \bbf \big(\bbs_{\tslice}, \bbp^{\bbphi}(\bbs_{\tslice}, \bblambda; \bbphi) \big) \Big]. \label{eq:augmented_lagrangian}
\end{aligned}
\end{equation}

An optimal state-augmented parametrization is defined to be a minimizer of the expectation of the Lagrangian w.r.t. a dual multiplier sampling distribution $q_{\bblambda}$, i.e.,
\begin{align}\label{eq:optimal_state_augmented_parametrization}
    \bbphi^\star \in \argmin_{\bbphi \in \bbPhi} \E_{\bblambda \sim q_{\bblambda}} \ccalL_{\bblambda}(\bbphi).
\end{align}

Given that the learning parameterization is sufficiently expressive (e.g., deep neural networks), the learned optimal state-augmented parameterization $\bbphi^\star$ approximates the Lagrangian minimizers $\bbtheta^{\dagger}(\bblambda) := \argmin_{\bbtheta \in \bbTheta} \ccalL_{\bblambda}(\bbtheta)$ for any $\bblambda \sim q_{\bblambda}$. 

The offline training of the state-augmented algorithm solves~\eqref{eq:optimal_state_augmented_parametrization}, which in practice can be accomplished by replacing the expectation with training batch averages and minimizing the empirical average Lagrangian using gradient-descent-based approaches. Given a batch of sequences of network states and augmenting dual multipliers $
\{ \bblambda_b, \{ \bbs_{b, \tslice} \}_{\tslice = 0}^{\Tslice-1} \}_{b =0 }^{B-1}$, the empirical Lagrangian is defined as
\begin{align}
\label{eq:lagrangian_empirical}
    \hat{\ccalL}(\bbphi_n) := \frac{1}{B} \sum_{b=0}^{B-1} \ccalL_{\bblambda_b}(\bbphi_n),
\end{align}
and the state-augmented model parameters are updated as
\begin{equation}
    \begin{aligned} \label{eq:state-augmented-gradient-descent}
        \bbphi_{n+1} &= \bbphi_{n} - \eta_{\bbphi} \nabla_{\bbphi} \hat{\ccalL}(\bbphi_n).
    \end{aligned}
\end{equation}

During the online execution phase, we derive the network-slicing decisions using the trained state-augmentation parametrization augmented with the current dual multipliers, and every ${\Tslice}_0$ slicing windows, dual multipliers are updated proportional to the constraint slacks as
\begin{equation}
\begin{aligned}\label{eq:dual_updates}
    \bblambda_{k+1} \!=\! \Big[ \bblambda_k + \frac{\eta_{\bblambda}}{{\Tslice}_0} \sum_{\tslice = k{\Tslice}_0}^{(k+1){\Tslice}_0-1} \bbf \big(\bbs_{\tslice}, \bbp^{\bbphi}(\bbs_{\tslice}, \bblambda_k; \bbphi^\star) \big) \Big]_{+}.
\end{aligned}
\end{equation}
A suitable choice of the update window hyperparameter $T_0$ depends on the variance of the constraint function and implies a trade-off between obtaining an unbiased, low variance estimate of the gradients and the number of iterations needed for the dual multipliers to converge to their near-optimal values. Similarly, given the varying nature of the physical metric each constraint pertains to (e.g., rate measured in megabits whereas latency measured in milliseconds), different dual step sizes can be used to update the different components of the dual multiplier vector. For clarity of presentation, we adopted the scalar notation to denote the dual step size throughout the paper and applied a suitable normalization of the constraint function in our experiments (see Section~\ref{subsec:experimental-setup}).

The training and test phases of the state-augmented algorithm are summarized in~Fig.~\ref{fig:train-and-test}. We note that under some mild technical assumptions, near-optimality and almost sure feasibility guarantees of \eqref{eq:DGA} extend to the state-augmented primal-dual algorithm. Moreover, although we do not consider a CRL problem, our proposed algorithm is amenable to tackling CRL problems. We refer the reader to \cite{calvo2021state} for a theoretical exposition on the state-augmented primal-dual algorithms.
\section{Experimental Results}\label{sec:experimental results}
\subsection{QoS Evaluation Metrics}
Recall that we operate over $\Tslice$ slicing windows indexed by $\tslice \in [0, \Tslice-1]$ and we further index any timestamp within each slicing window with $\tau \in [0, \tau_{\max})$. Given a FRU allocation $\bbp(\tslice)$ and associated channel state $\bbh^{\tslice}(\tau)$, we can define the (ergodic) average throughput $r_i(\tslice)$ over a slicing period $\tslice$ for each flow $i$ as follows. The instantaneous data rate with a current channel state $h$ and corresponding SNR is given by the Shannon rate $g(h) := \log(1 + h/\sigma^2)$. The full instantaneous rate for flow $i$ can be defined as
\begin{equation} \label{eq:instantaneous_rate}
    r^{\tslice}_i(\tau) = x_{i}(\tslice) \cdot g\left(h^{\tslice}_i(\tau) \right),
\end{equation}
where $x_i(\tslice) = \{ p_{m}(\tslice) \cond i \in \ccalU_m \}$ is the fraction of the total bandwidth $W$ assigned to the SLA category of flow $i$ and the average throughput is given by
\begin{equation}\label{eq:average_throughput}
    r_i(\tslice) = \E_{\tau \in [0, \tau_{\max}]} \left[ r^{\tslice}_i(\tau) \cdot \beta^{\tslice}_i(\tau) \right]
\end{equation}
where $\beta^{\tslice}_i(\tau)$ evaluates to $1$ whenever flow $i$ is scheduled to transmit and has nonempty packet buffer and $0$ otherwise. We utilize a uniform round-robin scheduler.
 
Let the set $\mathcal{J}^{\tslice}_i = \{j^{\tslice}_{i, 1},  j^{\tslice}_{i, 2}, \ldots, j^{\tslice}({i, N^{\tslice}_i}) \}$ denote the transmitted packets of flow $i$ during slicing window $\tslice$. For a packet $j \in \mathcal{J}^{\tslice}_i$, we can define the packet latency as
\begin{equation}\label{eq:packet_latency}
    \ell^{\tslice}_i(j) := \frac{\tau_j - \alpha_j}{P} + \frac{1}{r^{\tslice}_i(\tau_j) P},
\end{equation}
where $r^{\tslice}_i(\tau_j)$ is the instantaneous rate during transmission of packet $j$ as defined in~\eqref{eq:instantaneous_rate}, $P > 0$ is the fixed packet size, $\alpha_j$ and $\tau_j$ are the arrival and transmission times of packet $j$, respectively. Then the packet delay is defined as the sum of queueing delay $\tau_j - \alpha_j$ plus the actual transmission delay incurred. We finally define the latency for a given flow over a slicing window $\tslice$ as the maximum packet latency, i.e.,
\begin{equation} \label{eq:latency}
    \ell_i(\tslice) := \max_{j \in \mathcal{J}^{\tslice}_i} \ell^{\tslice}_i(j).
\end{equation}

\subsection{Experimental Setup}\label{subsec:experimental-setup}

\begin{figure*}[t!]
\centering
\begin{subfigure}[t]{\linewidth}
    \centering
    \includegraphics[width = \linewidth]{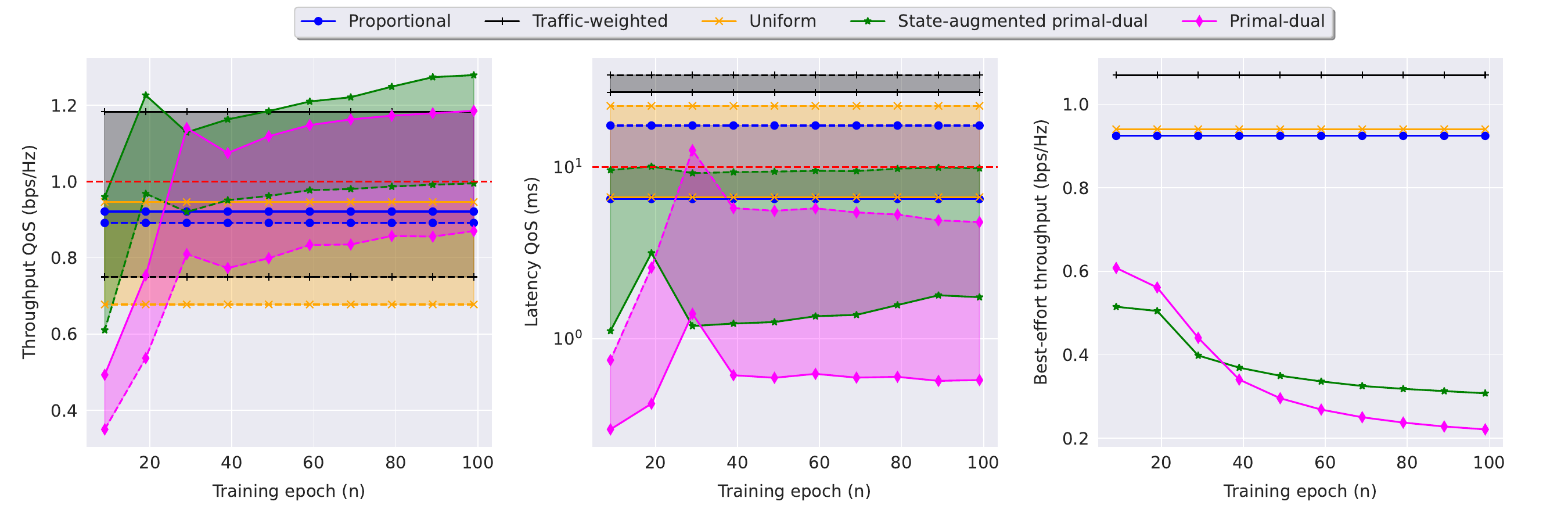}
    \label{fig:test_evolution}
\end{subfigure}\hfill%
\vspace{-1.5em}
\caption{Test evolutions of high-throughput (left), low-latency (middle) QoS constraints, best-effort throughput objective (right) for the proposed state-augmented primal-dual algorithm and the baselines. Objective and the constraints are averaged (solid lines) over $128$ test network realizations and the shadowed bands cover to the worst ($99^{\text{th}}$ percentile) realization (dashed lines). Red dashed lines indicate the respective QoS requirements.}
\label{fig:test_evolution}
\end{figure*}

The total bandwidth is $W = 20$ MHz. We fix the total number of flows to $|\ccalU| =20$ and randomly sample the (initial) number of H, L and B flows for each network configuration. Packet generations and traffic arrival for all flows follow a constant bit rate model, mean packet generation rates $\mu^{\tslice}_i$ are (uniform) randomly sampled and dynamic packet generation rates $\alpha(\mu^{\tslice}_i)$ vary across slicing windows following a random-walk with noise standard deviation of $0.5$ bps/Hz. The initial average data rates are sampled randomly $\mathrm{Unif}[1, 5]$ bps/Hz for high-throughput and best-effort flows and $\mathrm{Unif}[0.5, 1.5]$ bps/Hz for the low-latency slices. We adopt the Rayleigh fading channel model and the transmitter and access-point (AP) deployment approach of \cite{StateAugmented_RRM_GNN_naderializadeh_TSP2022}.

We set a minimum throughput requirement of $\rminht = 1.0$ bps/Hz for the high-throughput slices and a maximum tail packet latency requirement of $\lmaxll = 10$ ms for the low-latency slices. Best-effort slices have no QoS constraints. The network slicing problem with throughput and latency QoS metrics described by \eqref{eq:average_throughput} and \eqref{eq:latency}, respectively, is mapped to the constrained optimization problem of \eqref{eq:optimization_problem} with state-augmented parametrization as
\begin{equation}
\begin{aligned}\label{eq:state-augmented_optimization_problem_experiment}
    \max_{\bbphi \in \bbPhi} \;& \frac{1}{\Tslice} \sum_{\tslice = 0}^{\Tslice-1} \frac{1}{\vert \Ube \vert} \sum_{i \in \Ube} r_i \big(\bbs_{\tslice}, \bbp^{\bbphi}(\bbs_{\tslice}, \bblambda_{\lfloor \tslice / \Tslice_0 \rfloor}; \bbphi) \big) \\
    \mathrm{s.t.} \;& \frac{1}{\Tslice}\sum_{\tslice=0}^{\Tslice-1} \left[  \max_{i \in \Uht} \left( 1 - \frac{r_i \big( \bbs_{\tslice}, \bbp^{\bbphi}(\bbs_{\tslice}, \bblambda_{\lfloor \tslice / \Tslice_0 \rfloor}; \bbphi) \big)}{\rminht} \right) \right] \leq 0, \\
    &\frac{1}{\Tslice}\sum_{\tslice=0}^{\Tslice-1} \left[ \max_{i \in \Ull} \left( \frac{{\ell}_i \big(\bbs_{\tslice}, \bbp^{\bbphi}(\bbs_{\tslice}, \bblambda_{\lfloor \tslice / \Tslice_0 \rfloor}; \bbphi) \big)}{\lmaxll} - 1 \right) \right] \leq 0.
\end{aligned}
\end{equation}
The objective is to maximize the average throughput for best-effort slices. We formulate the QoS constraints as the ergodic average of the worst-case metrics over the respective SLAs to account for varying number of flows. We normalize the throughput and latency metrics so that the dynamic ranges of the two constraints are similar and choose a scalar step size for the dual updates. Following \eqref{eq:optimization_problem}, we make the dependence of the QoS metrics on the network state and the policy explicit.

Slicing decisions are made across $\Tslice = 50$ windows and the dual multipliers are updated every $\Tslice_0 = 2$ slicing windows during online execution of the state-augmented primal-dual algorithm. At each slicing window, all flows within the same SLA category take turns to transmit their packets to the AP via a round-robin scheduling algorithm, in parallel for all SLA categories. During transmission, the whole subband allocated to the respective slice is utilized by the flows and we assume no interband interference. Although the duration of each slicing window can span several minutes/hours in practice, we simulate and evaluate over slicing windows of duration $\tau_{\max} = 50$ ms for the sake of faster simulations.

For both primal-dual policies, we use multilayer perceptron (MLP) parametrizations with two hidden layers of width $64$ followed by a hidden layer of width $32$ and a final softmax layer. The inputs to the models are the concatenation of the network state vector $\bbs_{\tslice}$ and the dual multiplier vector $\bblambda_{\lfloor \tslice / \Tslice_0 \rfloor}$. The output $\bbp(\tslice)$ is a vector of the fraction of the total bandwidth allocated to the slices. The network state vector at slicing window $\tslice$ is the concatenation of the vector of fraction of flows in each slice and the average and total data rates across each slice, estimated from the previous slicing window.

We train both models over $128$ different network configuration for $100$ training epochs with a learning rate of $10^{-4}$. The augmenting dual multipliers for the state-augmented model are sampled uniformly from $[\mathbf{0}, \bblambda_{\max}] \times [\mathbf{0}, \bblambda_{\max}]$ with $\bblambda_{\max} = \mathbf{1}$ initially and $\bblambda_{\max}$ is updated at the end of each training epoch by running the online dual updates on a validation set of training networks. We test the trained models on $128$ network realizations. Dual step size for primal-dual training and online execution of the state-augmented algorithm are set to $\eta^{PD}_{\bblambda} = 0.1$ and $\eta_{\bblambda} = 1.0$, respectively. 

We consider the following traditional slicing baselines:
\begin{itemize}
    \item \emph{Uniform:} The total bandwidth available is split equally among all slices, i.e., $\bbp = (1/3, 1/3, 1/3)$ in our setup.
    \item \emph{Proportional:} The total bandwidth is sliced proportional to the number of active flows in each slice.
    \item \emph{Traffic-weighted (TW):} The total bandwidth is sliced proportional to the actual total traffic demand. 
\end{itemize}

\begin{figure*}[t!]
\centering 
\begin{subfigure}[t]{.66\linewidth}
    \centering
    \includegraphics[width = \linewidth]{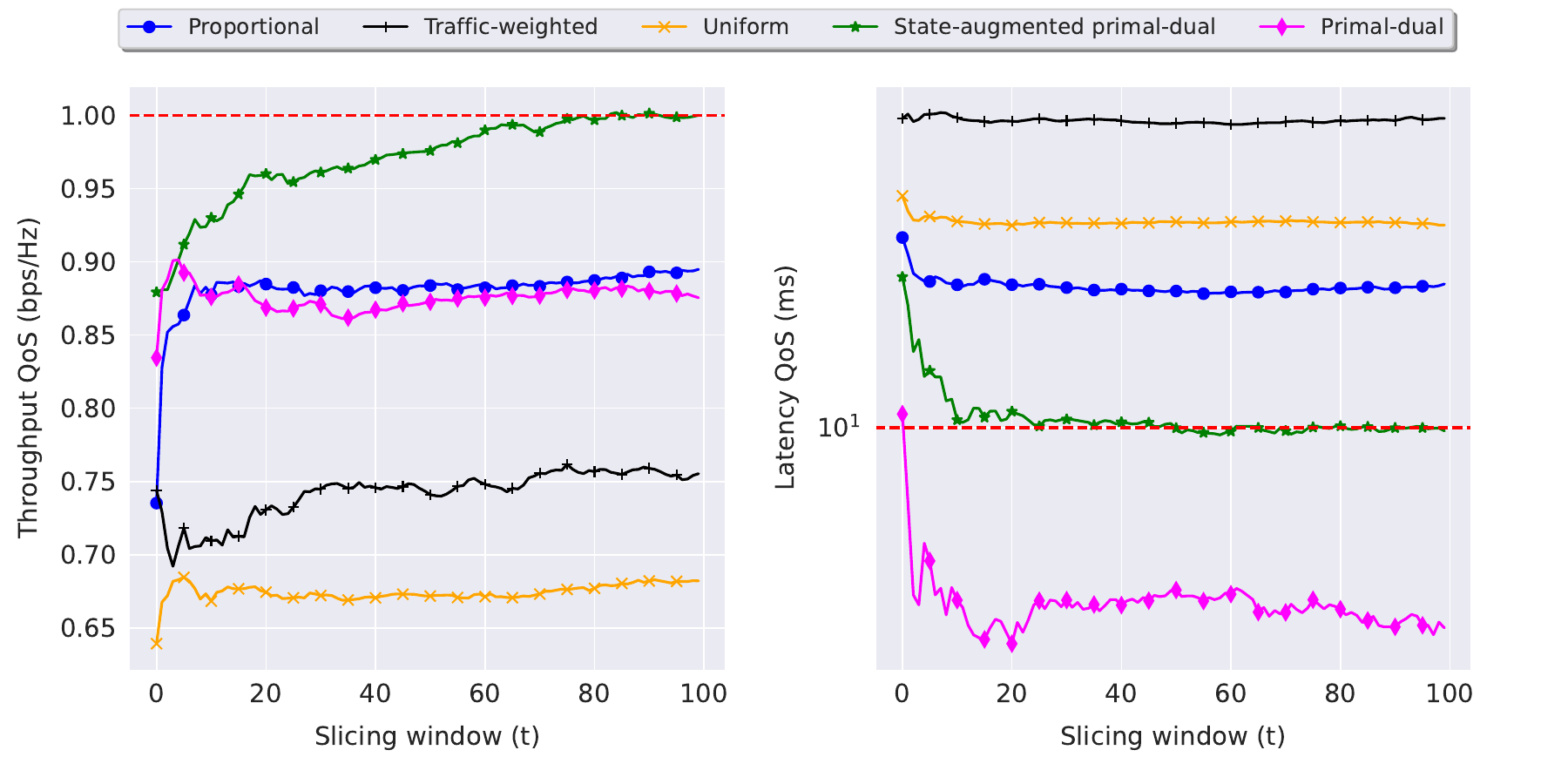}
    \vspace{-2em}
    \subcaption{QoS constraints}
\end{subfigure}\hfill%
\begin{subfigure}[t]{.33\linewidth}
    \centering
    \includegraphics[width = \linewidth]{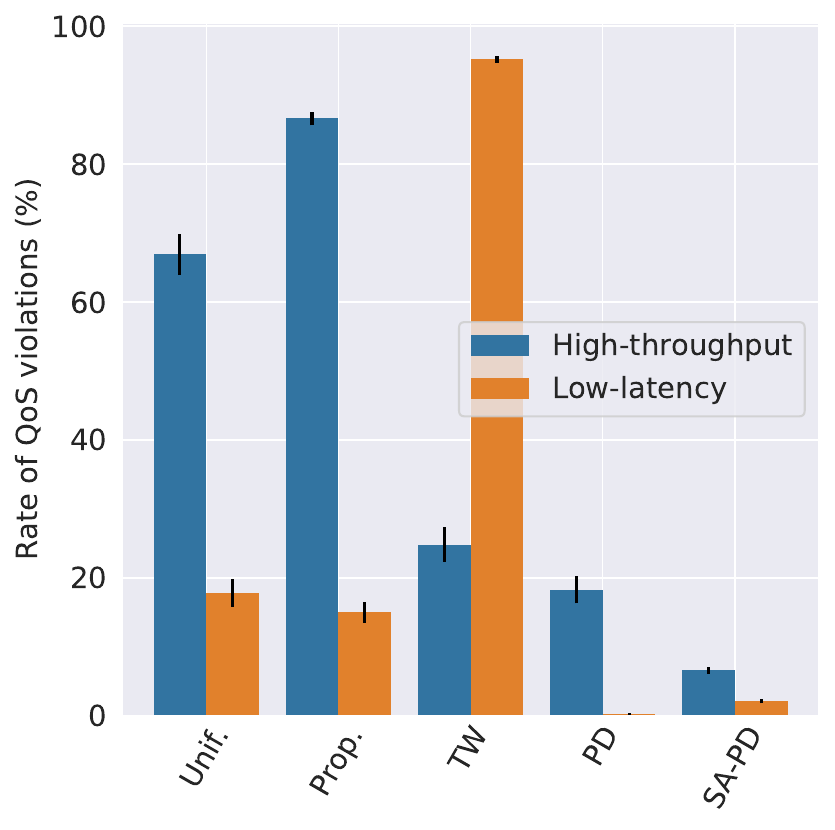}
    \vspace{-2em}
    \subcaption{Rate of QoS violations}
\end{subfigure}\hfill%
\caption{(a) Time evolution of cumulative ergodic averages of the QoS constraints for the worst realization and (b) bar plot of average rate of QoS violations.}
\label{fig:test_evolution_over_slices}
\end{figure*}

\begin{figure*}[t!]
\centering
    \begin{subfigure}[t]{.2\linewidth}
    \centering
    \includegraphics[height = \linewidth, width = \linewidth]{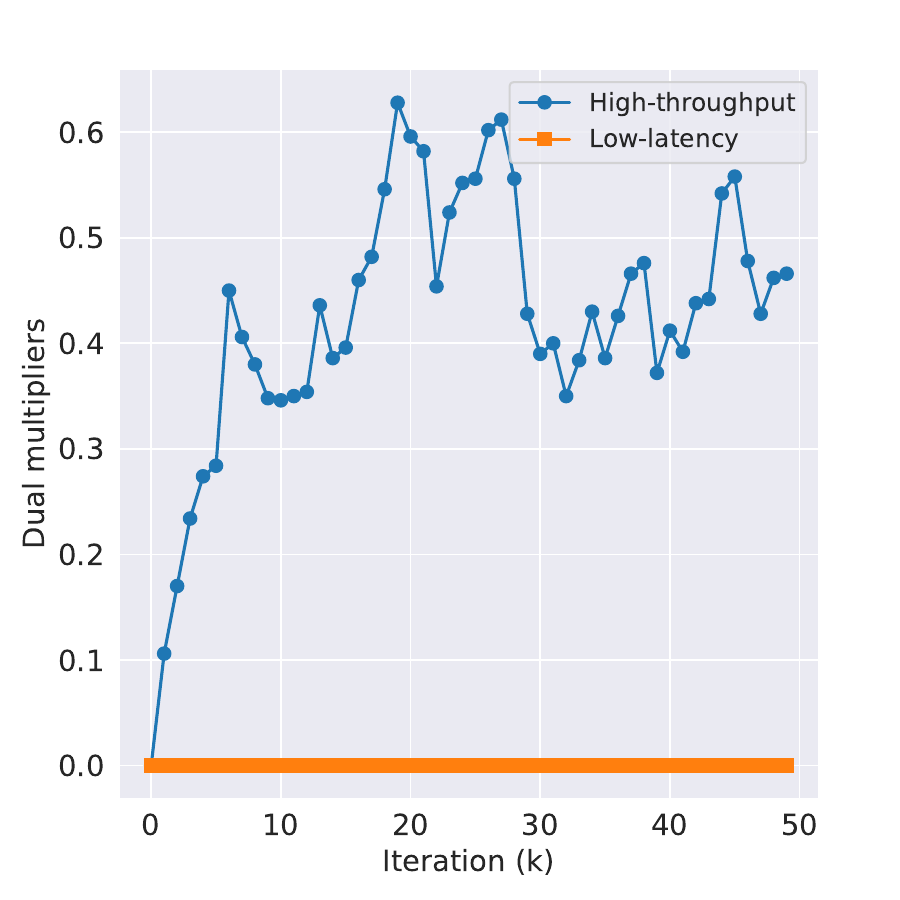}
    \subcaption{Dual multipliers}
    \end{subfigure}\hfill%
    \begin{subfigure}[t]{.4\linewidth}
    \centering
    \includegraphics[width = \linewidth, height = .5\linewidth]{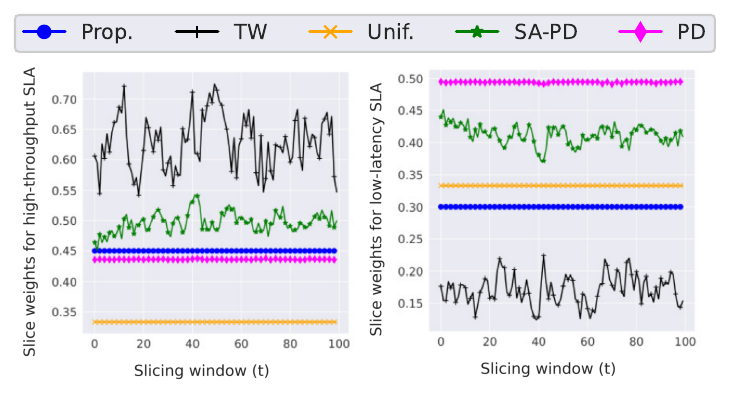}
    \subcaption{Slicing decisions}
    \end{subfigure}\hfill%
    \begin{subfigure}[t]{.4\linewidth}
    \centering
    \includegraphics[width = \linewidth, height = .5\linewidth]{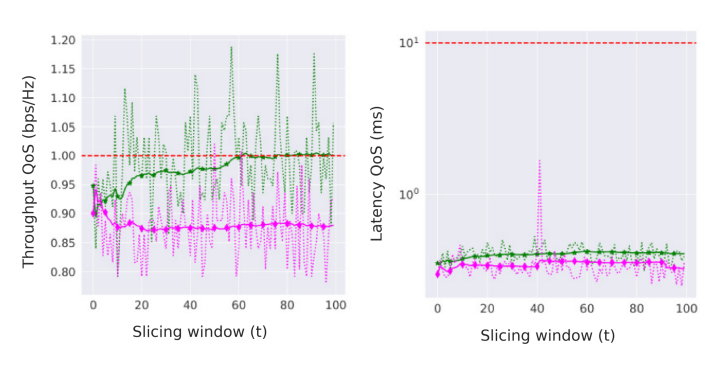}
    \subcaption{QoS constraints}
    \end{subfigure}\vfill%
\caption{Time evolution of (a) dual multipliers, (b) slicing decisions and (c) instantaneous (dotted) and average (solid) constraints over slicing windows for an example test network. Constraint averages are computed over a sliding window of length $\Tslice = 50$.}
\label{fig:test_evolution_constraint_metrics}
\end{figure*}%

\begin{table*}[t!]
\centering
\caption{Comparison of Rate of Instantaneous and Ergodic QoS Violations}
\label{tab:my-table}
\begin{tabular}{cc|clclclclcl|clclclclcl|}
\cline{3-22}
                                          &                    & \multicolumn{10}{c|}{\begin{tabular}[c]{@{}c@{}}High-throughput average instantaneous \\ and \textbf{ergodic} QoS violations (\%)\end{tabular}}                                                                                                                                                                                                                                             & \multicolumn{10}{c|}{\begin{tabular}[c]{@{}c@{}}Low-latency average instantaneous\\ and \textbf{ergodic} QoS violations (\%)\end{tabular}}                                                                                                                                                                                                                                               \\ \hline
\multicolumn{1}{|c|}{$r_{\min}$ (bps/Hz)} & $\ell_{\max}$ (ms) & \multicolumn{2}{c|}{Proportional}                                        & \multicolumn{2}{c|}{T.W.}                                                & \multicolumn{2}{c|}{Uniform}                                             & \multicolumn{2}{c|}{SA-PD}                                             & \multicolumn{2}{c|}{PD}                                                  & \multicolumn{2}{c|}{Proportional}                                        & \multicolumn{2}{c|}{T.W.}                                                & \multicolumn{2}{c|}{Uniform}                                             & \multicolumn{2}{c|}{SA-PD}                                             & \multicolumn{2}{c|}{PD}                                                \\ \hline
\multicolumn{1}{|c|}{0.7}                 & 5                  & \multicolumn{2}{c|}{\begin{tabular}[c]{@{}c@{}}0.3\\ \textbf{0.3}\end{tabular}}   & \multicolumn{2}{c|}{\begin{tabular}[c]{@{}c@{}}1.6\\ \textbf{0.1}\end{tabular}}   & \multicolumn{2}{c|}{\begin{tabular}[c]{@{}c@{}}7.6\\ \textbf{0.1}\end{tabular}}   & \multicolumn{2}{c|}{\begin{tabular}[c]{@{}c@{}}1.5\\ \textbf{0.9}\end{tabular}} & \multicolumn{2}{c|}{\begin{tabular}[c]{@{}c@{}}6.0\\ \textbf{4.7}\end{tabular}}   & \multicolumn{2}{c|}{\begin{tabular}[c]{@{}c@{}}27.5\\ \textbf{25.3}\end{tabular}} & \multicolumn{2}{c|}{\begin{tabular}[c]{@{}c@{}}98.5\\ \textbf{99.9}\end{tabular}} & \multicolumn{2}{c|}{\begin{tabular}[c]{@{}c@{}}35.6\\ \textbf{35.5}\end{tabular}} & \multicolumn{2}{c|}{\begin{tabular}[c]{@{}c@{}}0.4\\ \textbf{0.1}\end{tabular}} & \multicolumn{2}{c|}{\begin{tabular}[c]{@{}c@{}}0.3\\ \textbf{0.1}\end{tabular}} \\ \hline
\multicolumn{1}{|c|}{0.9}                 & 10                 & \multicolumn{2}{c|}{\begin{tabular}[c]{@{}c@{}}30.7\\ \textbf{2.2}\end{tabular}}  & \multicolumn{2}{c|}{\begin{tabular}[c]{@{}c@{}}13.5\\ \textbf{0.1}\end{tabular}}  & \multicolumn{2}{c|}{\begin{tabular}[c]{@{}c@{}}42.0\\ \textbf{56.2}\end{tabular}} & \multicolumn{2}{c|}{\begin{tabular}[c]{@{}c@{}}1.8\\ \textbf{0.1}\end{tabular}} & \multicolumn{2}{c|}{\begin{tabular}[c]{@{}c@{}}2.5\\ \textbf{2.0}\end{tabular}}   & \multicolumn{2}{c|}{\begin{tabular}[c]{@{}c@{}}14.9\\ \textbf{13.1}\end{tabular}} & \multicolumn{2}{c|}{\begin{tabular}[c]{@{}c@{}}95.4\\ \textbf{98.4}\end{tabular}} & \multicolumn{2}{c|}{\begin{tabular}[c]{@{}c@{}}18.1\\ \textbf{16.1}\end{tabular}} & \multicolumn{2}{c|}{\begin{tabular}[c]{@{}c@{}}4.0\\ \textbf{1.2}\end{tabular}} & \multicolumn{2}{c|}{\begin{tabular}[c]{@{}c@{}}1.3\\ \textbf{0.7}\end{tabular}} \\ \hline
\multicolumn{1}{|c|}{0.9}                 & 20                 & \multicolumn{2}{c|}{\begin{tabular}[c]{@{}c@{}}30.7\\ \textbf{2.5}\end{tabular}}  & \multicolumn{2}{c|}{\begin{tabular}[c]{@{}c@{}}13.5\\ \textbf{9.4}\end{tabular}}  & \multicolumn{2}{c|}{\begin{tabular}[c]{@{}c@{}}42.0\\ \textbf{56.2}\end{tabular}} & \multicolumn{2}{c|}{\begin{tabular}[c]{@{}c@{}}0.1\\ \textbf{0.1}\end{tabular}} & \multicolumn{2}{c|}{\begin{tabular}[c]{@{}c@{}}2.6\\ \textbf{1.6}\end{tabular}}   & \multicolumn{2}{c|}{\begin{tabular}[c]{@{}c@{}}0.5\\ \textbf{0.1}\end{tabular}}   & \multicolumn{2}{c|}{\begin{tabular}[c]{@{}c@{}}73.5\\ \textbf{76.3}\end{tabular}} & \multicolumn{2}{c|}{\begin{tabular}[c]{@{}c@{}}1.8\\ \textbf{1.0}\end{tabular}}   & \multicolumn{2}{c|}{\begin{tabular}[c]{@{}c@{}}3.8\\ \textbf{0.3}\end{tabular}} & \multicolumn{2}{c|}{\begin{tabular}[c]{@{}c@{}}1.2\\ \textbf{0.6}\end{tabular}} \\ \hline
\multicolumn{1}{|c|}{1.0}                 & 10                 & \multicolumn{2}{c|}{\begin{tabular}[c]{@{}c@{}}86.6\\ \textbf{99.9}\end{tabular}} & \multicolumn{2}{c|}{\begin{tabular}[c]{@{}c@{}}24.8\\ \textbf{18.0}\end{tabular}} & \multicolumn{2}{c|}{\begin{tabular}[c]{@{}c@{}}66.9\\ \textbf{56.2}\end{tabular}} & \multicolumn{2}{c|}{\begin{tabular}[c]{@{}c@{}}6.6\\ \textbf{3.8}\end{tabular}} & \multicolumn{2}{c|}{\begin{tabular}[c]{@{}c@{}}18.3\\ \textbf{18.0}\end{tabular}} & \multicolumn{2}{c|}{\begin{tabular}[c]{@{}c@{}}15.0\\ \textbf{13.0}\end{tabular}} & \multicolumn{2}{c|}{\begin{tabular}[c]{@{}c@{}}95.2\\ \textbf{98.3}\end{tabular}} & \multicolumn{2}{c|}{\begin{tabular}[c]{@{}c@{}}17.8\\ \textbf{15.8}\end{tabular}} & \multicolumn{2}{c|}{\begin{tabular}[c]{@{}c@{}}2.1\\ \textbf{0.2}\end{tabular}} & \multicolumn{2}{c|}{\begin{tabular}[c]{@{}c@{}}0.2\\ \textbf{0.1}\end{tabular}} \\ \hline
\end{tabular}
\end{table*}

\subsection{Simulation Results}
\subsubsection{Test Performance and Comparison with Baselines}
In Fig.~\ref{fig:test_evolution}, we present the test performance of the primal-dual (PD) and state-augmented primal-dual (SA-PD) algorithms against the baselines. The evolution of the test performances of the primal-dual algorithms with respect to the training epochs are plotted and we validate that the state-augmented algorithm meets the throughput and latency QoS requirements while none of the traditional baselines meet the constraints simultaneously. The primal-dual algorithm iterates, although feasible for an average network, do not converge to strictly feasible policies for all networks and constraints simultaneously unlike the state-augmented algorithm. Indeed, the worst throughput and latency metrics for the primal-dual algorithm (dashed purple line) oscillate around the QoS requirements. That the average objective value is lower for the primal-dual algorithms is a consequence of relatively strict QoS requirements. Ultimately, the best-effort slices have the lowest priority as the name implies.

\begin{figure}
\centering
    \begin{subfigure}[t]{.49\linewidth}
    \centering
    \includegraphics[width = \linewidth]{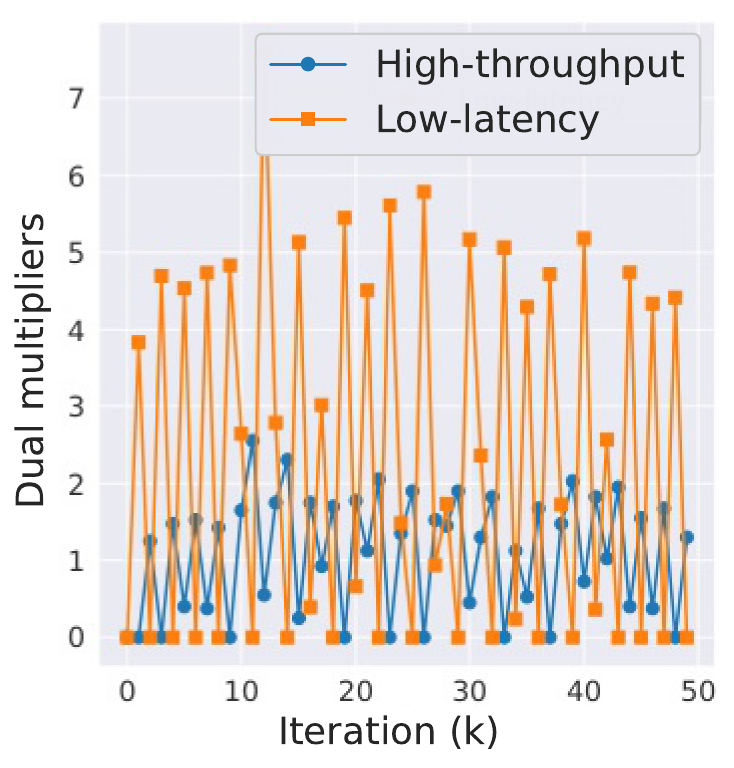}
    \end{subfigure}\hfill%
    \begin{subfigure}[t]{.51\linewidth}
    \centering
    \includegraphics[width = \linewidth]{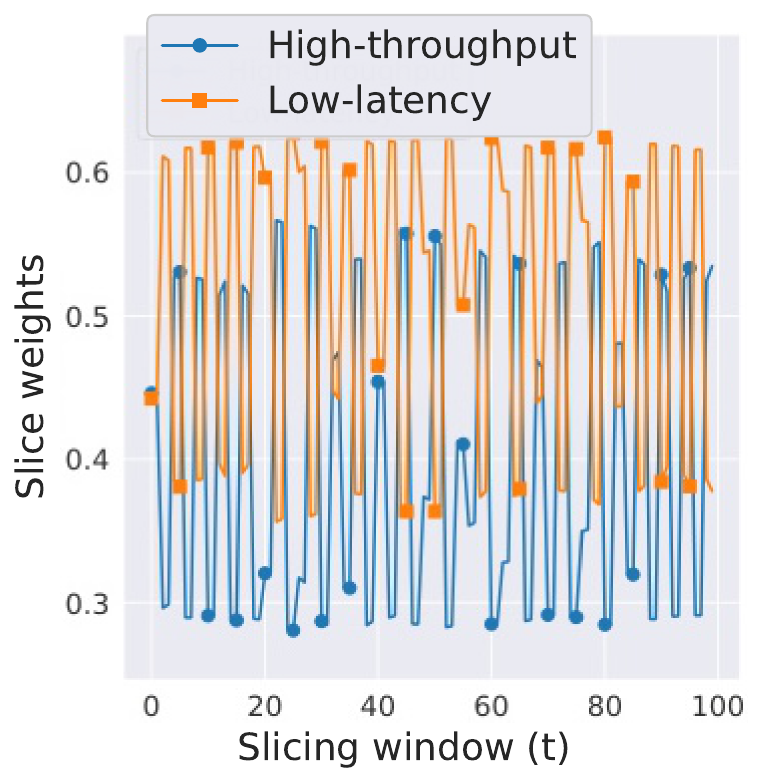}
    \end{subfigure}%
\caption{Example of policy switching with state-augmentation.}
\label{fig:policy_switching}
\end{figure}

\subsubsection{Online Execution of the State-Augmented Algorithm}
The ergodic feasibility of the state-augmented slicing policy is verified in Fig.~\ref{fig:test_evolution_over_slices}(a) where we plot the evolution of the QoS constraints for the worst network realization across slicing windows for fully-trained models. In Fig.~\ref{fig:test_evolution_over_slices}(b), we report the average rate of instantaneous QoS violations and the state-augmented algorithm achieves a low rate of instantaneous violation for both constraints on top of ergodic feasibility.

Fig.~\ref{fig:test_evolution_constraint_metrics} illustrates the online execution of the state-augmented algorithm for an example test network. As the throughput constraint is initially violated, the corresponding dual multipliers and slice weights allocated to high-throughput slices increase until the throughput constraint is met. 

\subsubsection{Policy Switching}
Fig.~\ref{fig:policy_switching} exemplifies the policy switching behavior of the state-augmented algorithm where alternating spikes in dual multipliers lead to switching between prioritizing high-throughput and low-latency slices to maintain the feasibility of both constraints. Neither of the policies that prioritize only either one of the two slices solve the constrained problem at hand. Finally, we report in Table~\ref{tab:my-table} the average rate of instantaneous and ergodic QoS violations for varying constraint level specifications. An ergodic throughput QoS violation occurs for a flow $i \in \Uht$ whenever $(1/\Tslice) \sum_{\tslice = 0}^{\Tslice-1} r_i(\tslice) < \rminht$ happens. Ergodic latency violations are defined similarly. Both primal-dual algorithms outperform the traditional baselines and the state-augmented algorithm attains comparable, if not better, ergodic QoS satisfaction.

\subsection{Practical Implications \& Discussion}
We hold the view that state-augmentation provides a principled alternative to direct policy randomization/averaging by learning a single parametrization to minimize the Lagrangian for all dual multipliers and running dual dynamics only during online execution. Accordingly, the dual dynamics sample different policies over time through policy-switching. Should a stochastic policy not be needed to solve the problem, the dual dynamics converge and the state-augmented algorithm is, although viable, not needed. Moreover, the Lagrangian minimizing policies do not depend on the constraint tolerances, only the optimal dual multipliers do. Given that the dual multipliers sampled during training cover a suitable range, a single state-augmented model could easily adapt to the changes (within a reasonable range) in the constraint specifications after deployment. With conventional primal-dual training, it would be necessary to either train different models for different operation regimes or condition the model on the constraint levels during training. 

One limitation of the state-augmented solution is that the choice of the uniform random sampling scheme for the dual multipliers is arbitrary and dependent on the application at hand. Second, we execute a suboptimal policy during the transient phase in the beginning before the dual multipliers reach their near-optimal values. Further research on near-optimal sampling and initialization of state-augmenting dual multipliers could significantly improve the proposed approach.

We presented network slicing as a timely and relevant application of state-augmentation. Many other wireless control and resource allocation problems of practical importance (e.g., multiple-access channels, power control, beamforming, routing and link scheduling etc.) are captured by the abstract problem formulation in \eqref{eq:optimization_problem} and can be tackled with a state-augmented primal-dual algorithm.

\section{Conclusion and Future Work}
We formulated (Wi-Fi) network slicing as a constrained learning problem. Through numerical simulations, we illustrated that a state-augmented algorithm that minimizes an expected Lagrangian offline and updates the dual multipliers online lends itself to a practical primal-dual algorithm with feasibility and optimality guarantees for network slicing. Extension to a state-augmented RL algorithm for network slicing, exploring the impact of imperfect CSI and end-to-end evaluation on a more realistic simulation platform are interesting future research directions. 

\bibliographystyle{IEEEbib}
\bibliography{refs.bib}

\begin{thebibliography}{10}

\bibitem{shen2020aiassisted}
Xuemin Shen, Jie Gao, Wen Wu, Kangjia Lyu, Mushu Li, Weihua Zhuang, Xu~Li, and Jaya Rao,
\newblock ``{AI}-assisted network-slicing based next-generation wireless networks,''
\newblock {\em IEEE Open Journal of Vehicular Technology}, vol. 1, pp. 45--66, 2020.

\bibitem{wu2022mobile6g}
Wen Wu, Conghao Zhou, Mushu Li, Huaqing Wu, Haibo Zhou, Ning Zhang, Xuemin~Sherman Shen, and Weihua Zhuang,
\newblock ``{AI}-native network slicing for {6G} networks,''
\newblock {\em IEEE Wireless Communications}, vol. 29, no. 1, pp. 96--103, 2022.

\bibitem{lemes2022tutorial}
Mario~Teixeira Lemes, Antonio~Marcos Alberti, Cristiano~Bonato Both, Antonio~Carlos De~Oliveira~Júnior, and Kleber~Vieira Cardoso,
\newblock ``A tutorial on trusted and untrusted non-{3GPP} accesses in {5G} systems—first steps toward a unified communications infrastructure,''
\newblock {\em IEEE Access}, vol. 10, pp. 116662--116685, 2022.

\bibitem{nerini20215g}
Matteo Nerini and David Palma,
\newblock ``{5G} network slicing for {Wi-Fi} networks,''
\newblock in {\em 2021 IFIP/IEEE International Symposium on Integrated Network Management (IM)}. IEEE, 2021, pp. 633--637.

\bibitem{networkslicing2018}
WB~Alliance,
\newblock ``Network slicing: Understanding {Wi-Fi} capabilities,''
\newblock Tech. {R}ep., March 2018.

\bibitem{zangooei2023reinforcement}
Mohammad Zangooei, Niloy Saha, Morteza Golkarifard, and Raouf Boutaba,
\newblock ``Reinforcement learning for radio resource management in {RAN} slicing: A survey,''
\newblock {\em IEEE Communications Magazine}, vol. 61, no. 2, pp. 118--124, 2023.

\bibitem{yang2023advancing}
Kun Yang, Shu ping Yeh, Menglei Zhang, Jerry Sydir, Jing Yang, and Cong Shen,
\newblock ``Advancing {RAN} slicing with offline reinforcement learning,'' 2023.

\bibitem{liu2020constrained}
Yongshuai Liu, Jiaxin Ding, and Xin Liu,
\newblock ``A constrained reinforcement learning based approach for network slicing,''
\newblock in {\em 2020 IEEE 28th International Conference on Network Protocols (ICNP)}. IEEE, 2020, pp. 1--6.

\bibitem{liu2021onslicing}
Qiang Liu, Nakjung Choi, and Tao Han,
\newblock ``Onslicing: online end-to-end network slicing with reinforcement learning,''
\newblock in {\em Proceedings of the 17th International Conference on emerging Networking Experiments and Technologies}, 2021, pp. 141--153.

\bibitem{agostini2022learning}
Patrick Agostini, Ehsan Tohidi, Martin Kasparick, and S{\l}awomir Sta{\'n}czak,
\newblock ``Learning constrained network slicing policies for industrial applications,''
\newblock in {\em 2022 IEEE International Conference on Communications Workshops (ICC Workshops)}. IEEE, 2022, pp. 1--6.

\bibitem{calvo2021state}
Miguel Calvo-Fullana, Santiago Paternain, Luiz Chamon, and Alejandro Ribeiro,
\newblock ``State augmented constrained reinforcement learning: Overcoming the limitations of learning with rewards,''
\newblock {\em IEEE Transactions on Automatic Control}, vol. PP, pp. 1--15, 01 2023.

\bibitem{StateAugmented_RRM_GNN_naderializadeh_TSP2022}
Navid NaderiAlizadeh, Mark Eisen, and Alejandro Ribeiro,
\newblock ``State-augmented learnable algorithms for resource management in wireless networks,''
\newblock {\em IEEE Transactions on Signal Processing}, 2022.

\bibitem{paternain2019zerodualitygap}
Santiago Paternain, Luiz Chamon, Miguel Calvo-Fullana, and Alejandro Ribeiro,
\newblock ``Constrained reinforcement learning has zero duality gap,''
\newblock in {\em Advances in Neural Information Processing Systems}, H.~Wallach, H.~Larochelle, A.~Beygelzimer, F.~d\textquotesingle Alch\'{e}-Buc, E.~Fox, and R.~Garnett, Eds., 2019, vol.~32.

\end{thebibliography}

\end{document}